\documentclass[aps,epsf,floats,preprint,epsfig, byrevtex, showpacs]{revtex4}

\usepackage{amsfonts}
\usepackage{amsmath}
\usepackage{amssymb}
\usepackage{graphicx}%

\newcommand{\beq}{\begin{equation}}
\newcommand{\eeq}{\end{equation}}
\newcommand{\beqn}{\begin{eqnarray}}
\newcommand{\eeqn}{\end{eqnarray}}
\newcommand{\bearr}{\begin{array}}
\newcommand{\enarr}{\end{array}}

\newcommand{\toref}[1]{\mbox{(\ref{#1})}}

\newcommand{\eps}{\varepsilon}
\begin{document}

\newcommand {\ee}[1] {\label{#1} \end{equation}}
\newcommand{\be}{\begin{equation}}

\preprint{ }

\title{On the relationship between directed percolation and 
the synchronization transition in spatially extended systems}
\author{ F. Ginelli$^{1,3}$, R. Livi$^{2,3}$, A. Politi$^{1,3}$ and A. Torcini$^{1,3}$}
\affiliation{$^1$Istituto Nazionale di Ottica Applicata,
           Largo E. Fermi 6, Firenze, I-50125 Italy \\ 
	   $^2$ Dipartimento di Fisica, Universit\`a di Firenze,
          Via Sansone 1 I-50019 Sesto Fiorentino, Italy \\
        $^3$Istituto Nazionale di Fisica della Materia, Unit\`a di Firenze
            Via Sansone 1 I-50019 Sesto Fiorentino, Italy}

\date{\today}
\widetext
\begin{abstract}
We study the nature of the synchronization transition in spatially extended
systems by discussing a simple stochastic model. An analytic argument is put
forward showing that, in the limit of discontinuous processes, the transition
belongs to the directed percolation (DP) universality class. The analysis is
complemented by a detailed investigation of the dependence of the first passage
time for the amplitude of the difference field on the adopted threshold. We
find the existence of a critical threshold separating the regime controlled by
linear mechanisms from that controlled by collective phenomena. As a result of
this analysis, we conclude that the synchronization transition belongs to the DP
class also in continuous models. The conclusions are supported by numerical
checks on coupled map lattices too.
\end{abstract} 


\maketitle

\section{Introduction}
\label{intro}

Synchronization in dynamical systems has recently become the subject of an
intensive research activity for various reasons that range from the application
to transmission of information, to the spontaenous onset of coherent behaviour
and also because because it is one of the mechanisms controlling the
degree of order present in a chaotic evolution. Most of the attention has
been, so far, focused on the behaviour of low--dimensional systems. As a
result of these investigations, several kinds of synchronizations have been
identified (the most important being phase and complete synchronization) and
the corresponding transition scenarios characterized \cite{PRK02}.

More recently, the interest has shifted towards high-dimensional chaos and,
specifically, on the behaviour of extended systems, a context in which an
overall picture is still lacking. In this paper, we devote our interest to
complete synchronization in lattice systems. This kind of synchronization has 
been introduced and studied into two different setups. In the former one, 
identical copies of a given system (with different initial internal states) 
converge to the same trajectory, when forced with the same random signal. This,
so-called stochastic, synchronization can occur only if the dynamics resulting
from the stochastic forcing becomes linearly stable, i.e. the maximum Lyapunov
exponent is negative \cite{FH92,MB94,P94,HF95,LZ98}. In the latter setup, two 
identical systems are coupled together: if the coupling strength is strong
enough, both eventually follow the same, chaotic, trajectory. This is the so
called chaotic synchronization. For it to be observed, it is sufficient that
the transverse Lyapunov exponent is negative \cite{P92}. Therefore, in
low--dimensional systems, synchronization  transition can always be reduced to
a linear stability problem.

On the other hand, recent numerical investigations \cite{BLT01,AP02} indicate
that the synchronization scenario in spatially extended dynamical systems 
exhibits more complex and interesting features. In fact, the addition of the 
spatial structure may turn the linear stability problem of low--dimensional 
systems into a nonequilibrium phase-transition problem.

In analogy to low--dimensional systems, various coupling schemes have been
already considered. For instance, ``stochastic synchronization'' has been
studied in Coupled Map Lattices (CML's), by adding the same spatio--temporal
noise $\xi(x,t)$ to different trajectories, $u_1(x,t)$ and $u_2(x,t)$, of 
the same system \cite{BLT01}, according to the following scheme
\beq
u_i(x,t+1) = f[u_i(x,t)+\nabla_{\eps}^2 u_i(x,t)] + \sigma \xi(x,t) \quad , 
\quad i=1,2 
\label{stosync}
\eeq
where
\beq
\nabla^2_{\eps}u(x,t) \equiv \frac{\eps}{2}u(x+1,t) + \frac{\eps}{2}u(x-1,t)
-\eps u(x,t) \quad,
\label{nablaeps}
\eeq
is the short--hand notation for the discretized Laplacian operator ($\eps$
plays the role of a diffusion constant) and $f[x]$ is a map of the unit
interval able to generate chaotic behaviour. Moreover, $\sigma$ is the
amplitude of the forcing term, $x$ is an integer index labeling the lattice
sites, $t$ is a discrete time variable and the noise term is assumed to be
bounded and $\delta$--correlated in space and time, i.e. $\langle
\xi(x,t)\xi(y,s) \rangle \propto \delta_{x,y} \, \delta_{t,s}$. Synchronization
is possible when the difference $w(x,t) = |u_1(x,t)-u_2(x,t)|$ between
simultaneous configurations of the two systems converges everywhere to zero.
The stability coefficient of the solution $w(x,t)=0$ is usually called the
Transverse Lyapunov Exponent (TLE). In the context of stochastic
synchronization, the evolution of a small $w(x,t)$ reduces to the tangent
dynamics of the single CML, so that the TLE coincides with the maximum Lyapunov
exponent of the noise-affected dynamics. Accordingly, synchronization can arise 
only when the stochastic forcing induces a negative maximum Lyapunov exponent.
This is possible if the
probability distribution of the state variable mostly concentrates in the 
region of the interval where the map acts as a contraction.

Alternatively, one can study the behavior of two directly coupled 
systems \cite{AP02}
\begin{eqnarray}
u_1(x,t+1) &=& (1-\sigma) f[u_1(x,t)+\nabla_{\eps}^2 u_1(x,t)] +
\sigma f[u_2(x,t)+\nabla_{\eps}^2 u_2(x,t)] ,\nonumber\\
u_2(x,t+1) &=& (1-\sigma) f[u_2(x,t)+\nabla_{\eps}^2 u_2(x,t)] +
\sigma f[u_1(x,t)+\nabla_{\eps}^2 u_1(x,t)]  \quad .
\label{chaosync}
\end{eqnarray}
At variance with the previous case, the coupling strength $\sigma$ modifies the 
evolution law of $w(x,t)$, by adding a stabilizing term, while it leaves 
unaffected the dynamics of the fully synchronized regime. Accordingly, the TLE
may become negative, while the maximum Lyapunov exponent, unchanged, remains
positive.

While the negativity of the TLE is always a necessary condition to observe
synchronization in spatially extended systems, for smooth enough dynamical
systems, it proves to be sufficient too. In fact, the study of stochastic and
chaotic synchronization, carried on in Refs.~\cite{BLT01} and \cite{AP02},
respectively, have shown that synchronization occurs as soon as the TLE
becomes negative and, correspondingly, the propagation velocity of
finite-amplitude perturbation vanishes. In particular, Ahlers and Pikovsky
\cite{AP02} argue that the dynamics of the coarse-grained absolute value 
$\tilde{w}$
of the difference field  is described by the following stochastic partial
differential equation,
\beq
\partial_t \tilde{w}(x,t) = D\nabla^2 \tilde{w}(x,t) + c_1 \tilde{w}(x,t) 
- c_3 \tilde{w}^3(x,t) + 
\tilde{w}(x,t) \eta(x,t) \quad,
\label{MNLangevin}
\eeq
where $D > 0$, $c_3 > 0$ 
and the Gaussian noise term $\eta$ is $\delta$--correlated in space and time,
i.e. $\langle \eta(x,t) \eta(y,s) \rangle \propto \delta(x-y) \delta(t-s)\,$.
This equation is formally equivalent to the mean--field equation of
the class of Multiplicative Noise (MN) nonequilibrium critical
phenomena \cite{GMT96}. By a Hopf--Cole transformation, 
$h(x,t)=-\ln \tilde{w}(x,t)$, the above equation can be 
transformed into \cite{foot1}
\beq
\partial_t h(x,t) = D\nabla^2 h(x,t) - D(\nabla h(x,t))^2 - (c_1-\frac{1}{2}) + 
c_3 e^{-2 h(x, t)} + \eta(x,t) 
\label{KPZW}
\eeq
describing the critical behaviour associated with the depinning transition of a
Kardar--Parisi--Zhang (KPZ) interface from a hard substrate. Numerical
analysis  confirms that the critical exponents evaluated for the two different
coupling schemes are both compatible with those predicted for the MN model.

On the other hand, it has been observed that in the presence of strong and
localized nonlinearities, the non-synchronized regime may coexist with a
negative TLE \cite{BLT01,AP02}. In this case, the transition does occur
when the propagation velocity of finite--amplitude perturbations vanishes, while
its critical properties turn out to belong to the class of Directed Percolation.
Such an equivalence  has been found by noticing that the fraction of 
non-synchronized sites (defined as those points where $|w(x)|$ is larger than some
small fixed threshold) is the appropriate order parameter corresponding to the
fraction of active sites in DP.

In this case, one cannot follow the same derivation as above, because even
close to the critical point, the evolution equation for $w(x,t)$ cannot be
linearized, since it is precisely the nonlinear effects which guarantee a
propagation of finite--amplitude perturbations in the presence of a negative
TLE.  It is worth recalling that in the formulation of Reggeon Field Theory,
the DP transition is described by the effective 
equation\cite{MD99,G95,H00} 
\beq
\partial_t \rho(x,t) = D\nabla^2 \rho (x,t) + c_1 \rho (x,t) - c_2 \rho ^2(x,t)  
+ \sqrt{\rho (x,t)} \eta(x,t) \quad,
\label{DPLangevin}
\eeq
where $\rho (x,t)$ is the density of active sites and $c_2 > 0$. 
Behind the similarity between
this and Eq. \toref{MNLangevin}, one should notice the crucial difference in the
noise amplitude: the square-root versus linear dependence on $\rho$ is indeed
responsible for turning the MN critical behaviour into a DP-like one. In this
paper, we plan to explain why the presence of a discontinuity (or a strong
nonlinearity) may lead to the effective equation (\ref{DPLangevin}). To this
aim, in Section 2 we introduce a simple Random Multiplier (RM) model as an
effective equation for the time evolution of the difference variable $w(x,t)$
for discontinuous and strongly nonlinear CML's. This model was originally
introduced in \cite{GLP02} to account for the mechanism of propagation of
information in stable chaotic systems. We analyze its phase  diagram, and we
also discuss how the synchronization transition may be modified when a true
discontinuity in the dynamics is changed into a strongly nonlinear continuous
mapping. The relation between the RM model and the DP mean--field  equation
(\ref{DPLangevin}) is analyzed in Section 3.
 
There is a further basic question that will be addressed here. All  microscopic
models that are known to exhibit a DP critical behaviour are  defined by
referring to discrete and finite state variables, such as the probabilistic
cellular automaton model proposed by Domany and Kinzel \cite{DK84}. In such
cases, the so--called ``absorbing state'' can be unambiguously identified. For
instance, in the cellular automaton of Ref.~\cite{DK84}, a sequence of ``$0$''s
can only change from its boundaries (this is the reason they are defined as
contact processes). In the context of synchronization, the dynamical variable
is continuous and the condition $w(x,t) = 0$ is never exactly fulfilled at any
finite time, even in a system of finite size. As a consequence, in numerical
experiments \cite{BLT01,AP02} one has to fix a small, but somehow  arbitrary,
threshold value, below which the trajectories are assumed to be  synchronized.
The same numerical simulations show that, independently of the  dynamical rule,
when the space average of $w(x,t)$ decreases below a  threshold value
${\mathcal O} (10^{-5})$ it does not grow again. However, one cannot {\sl a
priori} exclude that a large fluctuation of some  local multiplier drives the
system out of this weakly absorbing state. On the contrary, it looks plausible
to assume that in an infinite system such a large fluctuation occurs with
probability one. In Section 4  we tackle the problem of the existence of an
effective absorbing state even in the presence of a continuous state variable.
The study of the first passage  time required for the space average of the
difference variable $w(x,t)$ to go  through a series of decreasing thresholds
clarifies that, contrary to  intuition, it is possible to assign an effective
finite ``measure'' to the  synchronized, i.e. absorbing, state.
Finally, conclusions are drawn in Section 5.

\section{The Random Multiplier Model: definition and phase diagram}
\label{sec2}

In this section, we introduce the RM model with the aim of closely reproducing
the synchronization transition occurring in coupled piecewise linear maps of
the type,
\beqn
f(x)= \left\{  \begin{array}{lr} 
             x/\alpha_1  & 0\leq x < \alpha_1 \\ 
             1-(x-\alpha_1)/\alpha_2 &   \alpha_1 \leq x < \alpha_1+\alpha_2 \\
             (x - \alpha_1- \alpha_2)/\alpha_1 &  \alpha_1+\alpha_2 \leq x \leq 1  
\end{array} 
\right.\quad ,
\label{detmap}
\eeqn              
where $0<\alpha_1<1$ and $0<\alpha_2<1-\alpha_1$. For any $\alpha_2>0$, the map is continuous with a
highly expanding middle branch (when $\alpha_2 \ll 1$). In the limit $\alpha_2 =0$, $f(x)$
reduces to the discontinuous Bernoulli map with expansion factor $1/\alpha_1$.

In the {\it bidirectional synchronization} setup (\ref{chaosync}), the 
corresponding TLE is \cite{AP02}
\beq
\lambda_{\perp} =  \lambda_M  + \ln(1-2\sigma)  .
\label{tralyexp}
\eeq
where $\lambda_M$ is the maximum Lyapunov exponent of the single, uncoupled,
chain. Therefore, linear stability analysis indicates that a small deviation 
$w(x,t)= |u_1(x,t)-u_2(x,t)|$ is contracted when $\sigma > (1 - 1/\lambda_M)/2$. 
However, this is not the whole story even in the absence of multiplier
fluctuations, because whenever $u_1(x,t)$ and $u_2(x,t)$ fall
on different sides of the map-discontinuity, $w(x,t)$ becomes at once of order
1, being amplified by a factor close to $1/w(x,t)$. The probability of such
events depends on the probability density of the variables $u_i$: in the case
of a sufficiently smooth distribution across the discontinuity, the probability
is, to a leading order, proportional to $w$ itself \cite{foot2}. The same
qualitative behaviour does occur also for $\alpha_2 >0$,
except that now, when $w <\alpha_2$, the amplification factor cannot be larger
than $(1-2\sigma)\alpha_2$. Moreover, the probability of such amplifications does
no longer depend on $w$.

In the following, instead of determining the local dynamics of $w$ from the
actual evolution of $u_1(x,t)$ and $u_2(x,t)$, we prefer to write a self-contained
equation, where the occasional amplifications follow from a purely stochastic
dynamics that simulates the CML. More precisely, we introduce the model
\begin{equation}
v(x,t) = (1+\nabla^2_{\eps})w(x,t) \quad ,
\label{nabla}
\end{equation}
with
\begin{eqnarray}
\begin{array}{llr}
w(x,t+1) = \left\{
\begin{array}{ll}
1, & \mbox{w.p.} \quad p=a v(x,t) \\
a v(x,t), & \mbox{w.p.} \quad 1-p
\end{array}
\right., \quad 
\mbox{if $v(x,t) > \Delta$,} \\
\\
w(x,t+1)=\left\{
\begin{array}{ll}
v(x,t)/\Delta, & \mbox{w.p.} \quad p=a\Delta \\
a v(x,t), & \mbox{w.p.} \quad 1-p
\end{array}
\right., \quad
\mbox{if $v(x,t) \leq \Delta$,} 
\end{array}
\label{RMmodel}
\end{eqnarray}          
where $a$ and $\Delta$ replace $(1-2\sigma)/\alpha_1$ and $\alpha_2/(1-2\sigma)$, while
w.p. is a shorthand notation for ``with probability''. Only positive multipliers
are assumed in order to guarantee a positive defined $w(x,t)$ (simulations do
confirm that the sign does not play a relevant role). Finally, periodic boundary
conditions are assumed on a lattice of size $L$.

\begin{figure}[tcb]
\centering
\includegraphics*[width=8cm, angle=0]{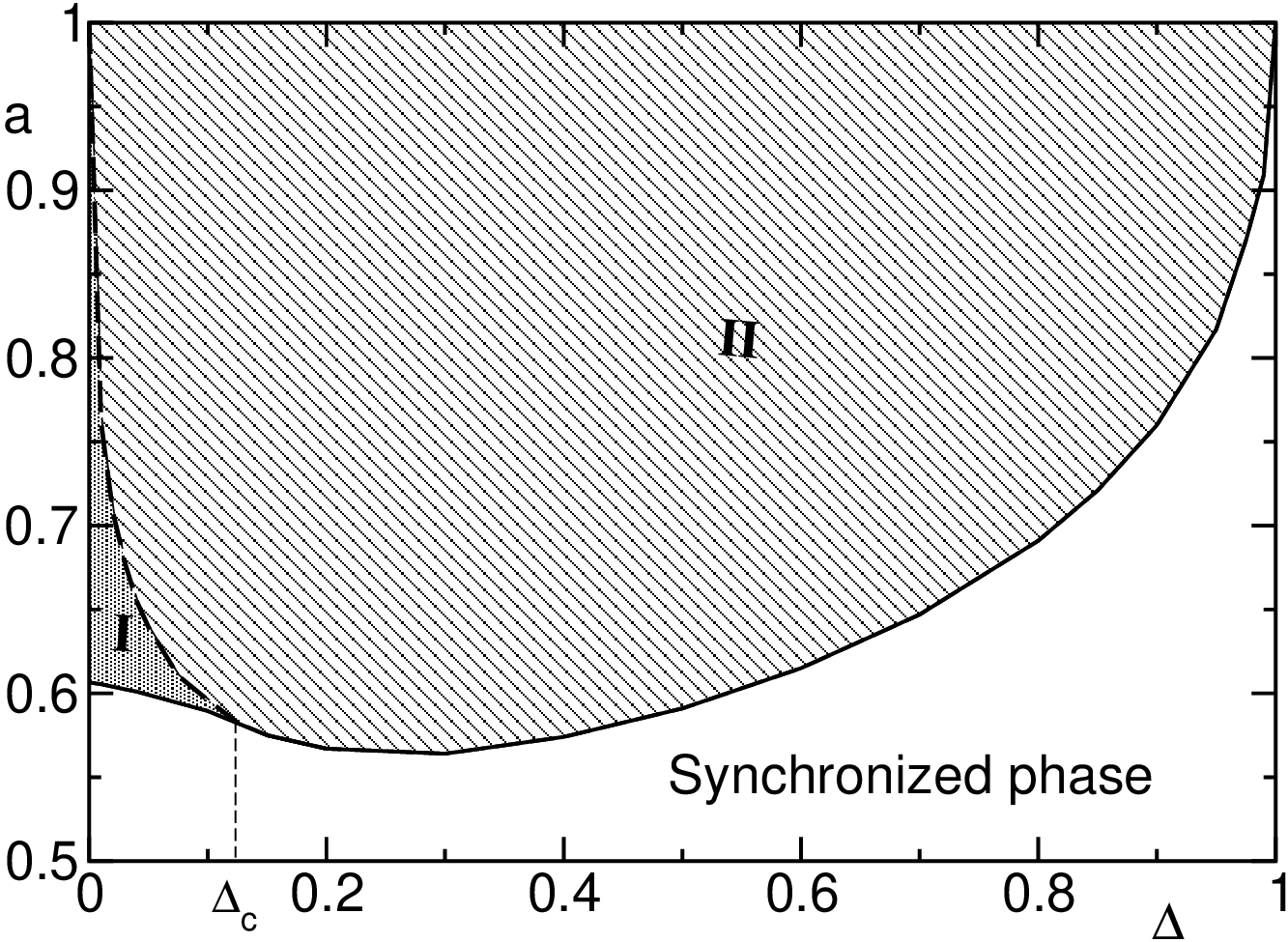}  
\caption{Phase diagram of the RM model. In the two shaded regions ( $I$ and 
$II$) the synchronized phase is unstable; in fact, above the solid line 
$v_F >0$. Along the dashed line, the TLE changes sign, so that region $I$ 
corresponds to the linearly stable, but nonlinearly unstable regime, while
$II$ corresponds the linearly unstable region. Above $\Delta=\Delta_c$,
the TLE vanishes together with $v_F$.}
\label{RMPhase} 
\end{figure} 

The advantage of playing with this model is that it explicitly avoids the 
possibly subtle correlation that may be generated during the deterministic
evolution of the CML and thereby spoiling the asymptotic behaviour of the 
observables we are interested in. Besides the probabilistic, rather than 
deterministic, choice of the amplification factor, the only other difference 
between the stochastic model (\ref{RMmodel}) and the original set of two 
coupled CMLs is the distribution of the amplification factors that is 
dichotomic in the former case. We see no reason why this difference should 
affect the transition scenario.

Moreover, in order to maximize propagation effects (that are responsible for
the propagation of finite-size perturbations) we shall restrict to the case 
$\eps = 2/3$ (the so called ``democratic'' coupling). Some rough numerical 
analyses do not, indeed, reveal qualitative changes when $\eps$ is varied 
around $2/3$. 

The most general way of testing the stability of the synchronized phase is by
monitoring the evolution of a droplet of the unsynchronized phase. By denoting 
with $N(t)$ the droplet size, i.e. the number of unsynchronized sites, at time
$t$, the propagation velocity can be defined as
\beq
v_F \equiv \lim_{t \rightarrow \infty} \frac{N(t)-N(0)}{2t} \quad .
\label{vF}
\eeq
A negative TLE (the maximum Lyapunov exponent of model (\ref{RMmodel})) implies
that any infinitesimal perturbation does decay. In spite of this linear 
stability, in Ref.~\cite{GLP02} it has been shown that $v_F$ can be positive, 
implying that the unsynchronized phase sustains itself and invades 
the synchronized one.
By performing detailed simulations for different values of the parameters $a$
and $\Delta$, we have been able to construct the phase diagram plotted in
Fig.~\ref{RMPhase}. The solid line, along which $v_F=0$, separates the 
synchronized from the unsynchronized phase (shaded region). The dashed line,
along which the TLE is equal to 0, splits the unsynchronized phase into a 
linearly stable ($I$) and unstable ($II$) region. In the former
one (ending approximately at $\Delta = \Delta_c \approx 0.15$), the nonlinear 
amplification mechanism prevails over the linear contraction induced by the 
negative TLE. Above $\Delta_c$, the TLE changes sign exactly where $v_F$ 
vanishes too.

\begin{figure}[tcb]
\centering
\includegraphics[width=14cm, angle=0]{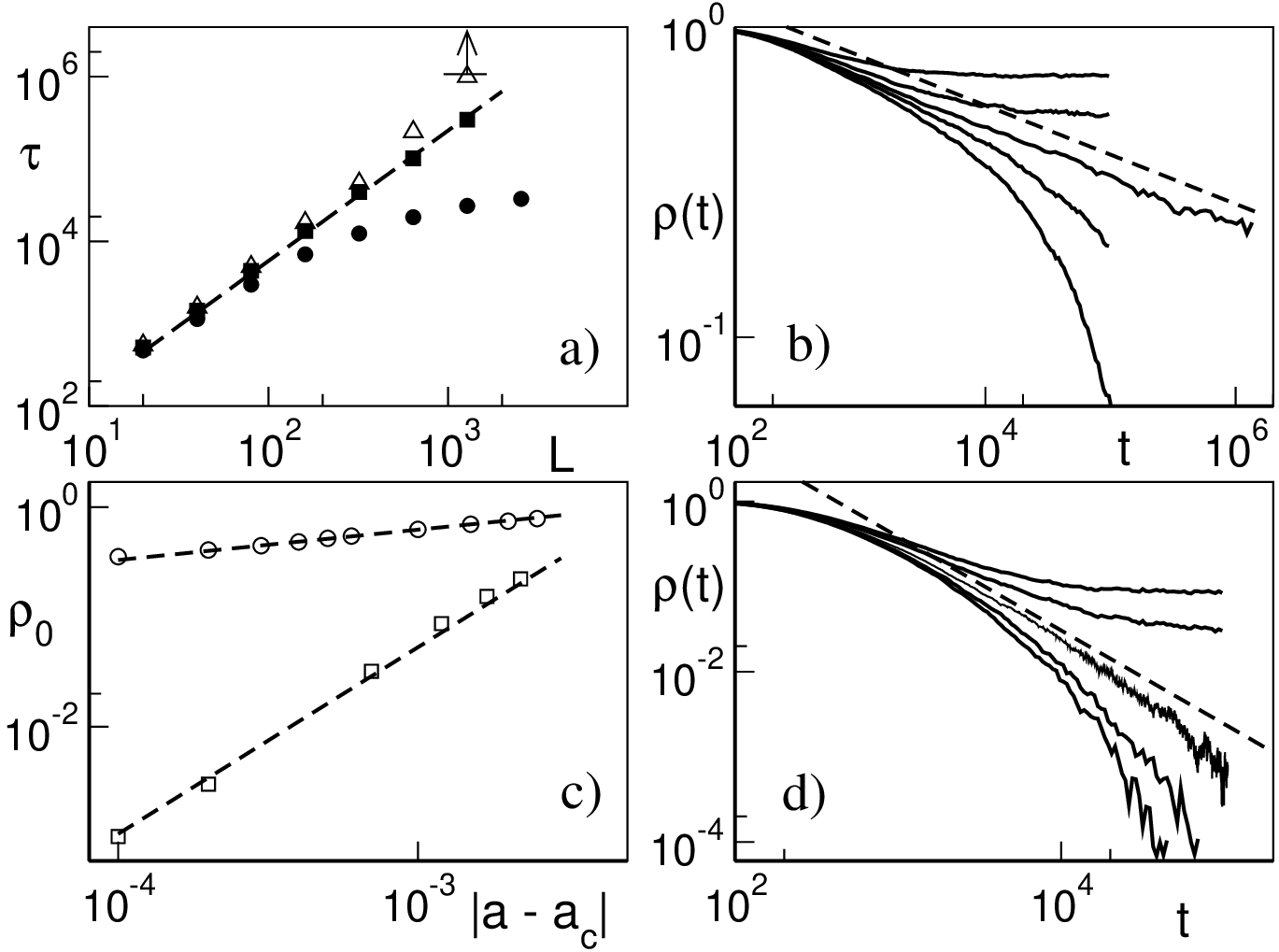}  
\caption{Power law scaling behaviour in the RM model. In all graphs, the dashed
lines correspond to the expected scaling behaviour (DP in a), b) and c); MN in
d)). a) Absorption time as a
function of system size for $\Delta = 0$. Triangles corresponds to $a=0.6070$,
squares to $a=0.6063$ and circles to $a=0.6051$. b) Density of unsynchronized
sites as a function of time for $\Delta=0.1$. The five solid lines correspond
to (from top to bottom) $a=0.591$, $a=0.59$, $a=0.58955$, $a=0.5893$ and
$a=0.5890$, respectively. c) Asymptotic density of unsynchronized sites as a
function of the distance from criticality. Circles corresponds to $\Delta=0.1$
and squares to $\Delta = 0.2$. d) Density of unsynchronized sites as a function
of time for $\Delta=0.2$. The five solid lines correspond to (from top to
bottom) $a=0.568$, $a=0.5675$, $a=0.5668$, $a=0.5664$ and $a=0.5662$. All the
graphs are plotted in a doubly logarithmic scales.} 
\label{RMcritical} 
\end{figure} 
Numerical analysis of stochastic synchronization in CML \cite{BLT01} suggests 
that when the TLE vanishes together with $v_F$, the critical properties of the 
synchronization transition are those of the MN class, while the transition is
DP-like whenever $v_F$ only vanishes (the TLE remaining negative).

Before entering a quantitative discussion about the nature of the transition 
in the present model, it is worth noticing a difference between the regimes 
$I$ and $II$. Linear instability in $II$ ensures that any finite perturbation 
of a synchronized state remains finite forever independently of the chain 
length. On the other hand, in $I$, a finite perturbation eventually dies
in a finite chain. The reason why the synchronized regime can nevertheless 
be considered unstable is that the average life time of the perturbation 
diverges exponentially with the chain length. This is a typical property of
systems in DP universality class and it can be traced back to the peculiar 
nature of the ``square root'' noise amplitude in 
Eq.~\toref{DPLangevin} \cite{M98}.

A preliminary numerical analysis of the critical properties of the RM model for
$\Delta = 0$ and $\Delta = 0.01$ has been already published in \cite{GLP02}.
Here we both perform more accurate simulations and extend the previous study to
larger values ($\Delta = 0.1$, 0.2) in order to find a signature of the change
of critical behavior. In all cases, $a$ is chosen to be the control parameter,
while the averaged (over different noise realizations) 
density $\rho(t)$ of unsynchronized sites will be the order parameter. 
The definition of $\rho$ requires to fix a small threshold 
$W$ to discriminate between synchronized ($w(x,t) < W$) and 
unsynchronized ($w(x,t) > W$) sites. In principle, $\rho(t)$ depends on 
$W$, both because the perturbation reaches different thresholds at 
different times and resurgencies can occur. Numerical analysis, however, 
indicates that, in practice, if $W$ is chosen on the order or smaller
than $10^{-5}$ no appreciable differences are observed. We shall come back to 
this problem in Sec. 4, to provide a more sound justification for the adopted 
procedure.

In order to test the relationship between synchronization transition and the DP 
critical phenomenon, we have investigated the scaling behaviour in the vicinity
of the transition. In DP it is known that, at criticality, the dependence of 
the density $\rho(t)$ on $t$ and $L$ is described by the scaling relation
\cite{H00}
\beq
\rho(t) = L^{-\delta z} g\left(\frac{t}{L^z}\right)\quad,
\label{scale1a}
\eeq
where $z$ is the so-called dynamical exponent accounting for the dependence of
the average time $\tau$ needed for $\rho$ to vanish with the system size
$L$ \cite{ND88},
\beq
\tau \sim L^z
\quad\quad\quad a=a_c \quad.
\label{tauscaling}
\eeq
Since for small $\theta=t/L^z$, the scaling function behaves as 
$g(\theta) \sim \theta^{-\delta}$, the exponent $\delta$ turns out to describe
the power-law decay of $\rho(t)$ 
\beq
\rho(t) \sim t^{-\delta}
\quad\quad\quad a=a_c \quad.
\label{deltascaling}
\eeq
Finally, the exponent $\beta$ characterizes the scaling behaviour of the
saturated density of active sites $\rho_0(t)$ as a function of
the distance from the critical value,
\beq
\rho_0\sim (a-a_c)^{\beta}
\quad\quad \quad a>a_c \quad.
\label{betascaling}
\eeq In analogy with usual nonequilibrium phase transitions, $z$, $\delta$, and
$\beta$ are expected to characterize all critical properties of the
synchronization transition as well. In fact, simple dimensional arguments show
that the exponents ruling the power law divergence exhibited by space- and
time-correlation functions (while approaching the critical point) are linked to
the previous ones by the standard relations
\beq
\nu_{\parallel} = \frac{\beta}{\delta} \quad, \quad\quad \nu_{\perp} = 
\frac{\nu_{\parallel}}{z} \quad.
\eeq 
Some of the scaling behaviors have been plotted in Fig.~\ref{RMcritical} to show
the quality of the results, while a complete summary of the exponents are
reported in Table~\ref{RMtable}, together with the best known estimates for the
DP \cite{J96} and the MN \cite{TGM97} class. 

The dynamical exponent has been estimated by averaging the behaviour of
relatively small systems (from $L=2^5$ to $L=2^{10}$)  over a large number of
noise realizations (of order $10^3$). In order to minimize finite-size effects,
the exponents $\delta$ and $\beta$ have been estimated from the time evolution
of a single  system of size $L=2^{20}$, relying on the large size to reduce
statistical fluctuations.  In the MN context we have not been able to
estimate $z$ through the measure of the average synchronization time,  but we
verified, through finite size-scaling \toref{scale1a}, that the value
of the dynamical exponent is compatible with the theoretical prediction.

Interestingly, similar results are obtained by adopting a different order
parameter, i.e. the space averaged difference variable  
$w(t) = \langle w(x,t) \rangle_x$. Also in this case, both $\langle w(t)\rangle$ 
(where $\langle \cdot \rangle$ denotes ensemble average) and the
absorption time $\tau_1$, defined as the average time required for $w(t)$ to become
smaller than some threshold $W$, are found to follow the same critical
scaling laws. The application of coarse-graining suggests that the space
average is the ``natural'' order parameter in the context of both equilibrium
and nonequilibrium critical transitions.

\begin{table}[t]
\centering
\begin{tabular}{||c|c|c|c|c|c|c||} \hline
&$\Delta = 0$ & $\Delta = 0.01$ & $\Delta =0.1$ & $\Delta=0.2$ & DP & MN \\ \hline
$z$ & $1.56\pm 0.06$ & $1.58\pm 0.02$ & $1.54 \pm 0.06$  & $1.5$ * & $1.580745 
\pm 10^{-6}$  & $1.53 \pm 0.07$ \\ \hline
$\delta$ & $0.155 \pm 0.005$ & $0.15 \pm 0.01$ &  $0.159 \pm 0.002$ & $1.2 
\pm 0.1$
& $0.159464 \pm 6 \cdot 10^{-6}$ & $1.10 \pm 0.05$\\ \hline
$\beta$ & $0.24 \pm 0.02$ & $0.27 \pm 0.01$  & $0.27 \pm 0.01 $ & $1.8 \pm 0.1$
 & $0.276486 \pm 6 \cdot 10^{-6} $ & $1.70 \pm 0.05$ \\ \hline
$a_c$ & $0.6063\ldots$  & $0.605\ldots$  &$0.5895 \ldots$  &  $0.5668\ldots$ &
 & \\ \hline
\end{tabular}
\caption{Numerical results concerning the $z$ the $\delta$ and the $\beta$
exponent of the RM model are compared with the best available estimates.
Values for $a_c$ indicate our best estimates of the critical point. Errors
have been estimated as the maximum deviation from linearity in the log-log
plot that it is  used to extract the scaling law.
* Indicates a value compatible with the theoretically predicted one, $z=1.5$
(see text).}
\label{RMtable}
\end{table}

\section{From the RM model to the DP field equation}
\label{sec3}
In this section we investigate the connection between the RM model and the DP
field equation \toref{DPLangevin}. Let us first consider the simple case 
$\Delta=0$ that corresponds to a discontinuous but otherwise uniformly
contracting local map. Eq.~\toref{RMmodel} can be recasted as  
\begin{equation} 
w(x,t+1) = 2a v(x,t) - a^2 v^2(x,t) + g(v) \xi'(x,t) \quad,
\label{cont_stoch3} 
\end{equation}
where $\xi'(x,t)$ is a zero-average $\delta$-correlated noise with unit
variance. In fact,
\beq
\xi'(x,t) = \frac{1}{g(v)}[\xi_{v}(x,t)-\langle \xi_{v} \rangle] \quad, 
\eeq
where
\beq
\xi_v(x,t) = \left\{ 
\begin{array}{ll} 
1-v(x,t), & \mbox{w.p.} \quad p=a v(x,t) \\ 
(a-1)v(x,t), & \mbox{w.p.} \quad 1-p ,
\end{array}  
\right. \quad,
\label{2noise}
\eeq
and 
\begin{eqnarray}
 \langle \xi_v \rangle &=& (2a-1) v - a^2 v^2 \\
 g^2(v) &=& \langle \xi_v^2\rangle -\langle \xi_v\rangle^2  = a  - 
 3 a^2 v^2 +  3 a^3 v^3
- 5 a^4 v^4 \quad .
\label{mv}
\end{eqnarray}
If we now introduce the coarse grained variable 
$\rho(x,t)= \overline{w(x,t)}$ (where the bar denotes an average over a
suitable space-time cell), we have that
$\overline{v(x,t)} = \rho(x,t) + \frac{\eps}{2}\nabla^2 \rho(x,t)$
so that Eq.~(\ref{cont_stoch3}) yields,
\beq
\begin{array}{ll}
\partial_t \rho(x, t)=& 
a \eps \nabla^2 \rho(x, t) + (2a-1) \rho (x,t) - a^2 \rho ^2(x,t) 
+ a^2 \eps \rho(x,t) \nabla^2 \rho(x, t) + \\
& \frac{(a \eps)^2}{4} \left[\nabla^2 \rho(x,t)\right]^2
+ g\left(\rho(x,t) + \frac{\eps}{2} \nabla^2 \rho(x, t)\right) \eta(x, t) 
\quad.
\end{array}
\label{DP-a} 
\eeq
where, according to the central limit theorem \cite{KTH85}, the coarse grained
noise term $\eta(x, t)$ is Gaussian and $\delta$-correlated in time and space.
According to standard renormalization-group arguments, \cite{H00,MD99,G95} the
terms of order $(\nabla^2 \rho)^2$ and $\rho \nabla^2 \rho$ can be shown to be
irrelevant, as well as the terms of order higher than or equal to $\rho$ and
$\sqrt{\nabla^2 \rho}$ appearing in the noise amplitude  $ g(\rho +
\frac{\eps}{2} \nabla^2 \rho)$. 

From the definition (\ref{mv}) of $g$ and after discarding the irrelevant terms,
the above equation reduces to
\beq
\partial_t \rho(x, t)= 
a \eps \nabla^2 \rho(x, t) + (2a-1) \rho (x,t) - a^2 \rho ^2(x,t) 
+\sqrt{a \rho(x,t)} \eta(x, t) .
\label{DP-b} 
\eeq
which is nothing but Eq. \toref{DPLangevin}, thus proving that the synchronization
transition in discontinuous CML's can be traced back to a DP nonequilibrium
phase transition. 

Let us now turn our attention to the more general case $\Delta >0$, which corresponds 
to a continuous local
mapping. According to Eq.~\toref{RMmodel}, we have now to deal with two different
kinds of noise, depending whether $v(x,t) > \Delta$ or $v(x,t) \leq \Delta$. 
By repeating the same formal derivation as in the previous case, 
we find that Eq.~\toref{DP-b} still holds when $\rho >\Delta$, while for
$\rho(x,t) < \Delta$ it must be replaced by the equation 
\beq
\partial_t \rho(x, t)= 
\frac{a\eps}{2}\left(2 - a\Delta\right) \nabla^2 \rho(x,t) + (2a-1-a^2\Delta) \rho(x,t) 
+ h(w) \eta(x, t) \quad  , 
\label{cont_stoch5} 
\eeq
where
\beq
h(w) = \rho(x,t) \sqrt{\frac{a}{\Delta} - 3a^2 + (2+5\Delta)a^3 - 
 (3+2\Delta )\Delta a^4 + (1-\Delta^2)\Delta a^5} \quad. 
\label{hfunc}
\eeq
Accordingly, in Eq.~\toref{cont_stoch5}, the noise amplitude is proportional to
the field itself, so that one should be led to the naive conclusion that the DP
critical behaviour is destroyed as soon as $\Delta$ is finite, or,
equivalently, that {\it any} CML system characterized by a continuous local
mapping cannot exhibit a DP-like synchronization transition. However, the
simulations described in the previous section suggest that DP-like transition 
can still been found for small but finite values of $\Delta$. 
In the next sections we shall present theoretical arguments supporting 
such numerical findings. \\

\section{First passage times}
\label{sec4}

In this section we clarify the problem of how and when it is possible
to observe a DP-like scenario in models like the RM one, with no clearly
identifiable absorbing state. As already noted in \cite{GLP02}, in any
finite system (of length $L$) there always exists a finite probability for a
generic configuration to be contracted forever, i.e. absorbed.
A lower bound to such a probability is (in the discontinuous case), 
\beq 
P = \left[ \prod_{n=1}^{\infty} (1-w_{M}a^n)\right]^L \quad,
\label{p1}
\eeq 
where 
\beq 
w_{M} = \max_x{w(x,0)} \quad.
\label{umax} 
\eeq 
However, since the null state, $w(x,t) = 0$, is reached in an infinite time,
this configuration cannot be attained with perfect accuracy in numerical
simulations and one is, in fact, obliged to fix a small but finite threshold.

The best way we have found to characterize the dependence of the perturbation
evolution on its size is through an indicator closely related to the 
finite-size Lyapunov exponent (FSLE) introduced in Ref.~\cite{ABCPV96}. 
With reference
to a perturbation initially set equal to 1 ($w(x,0) = 1 ,\quad x=1 \ldots L$),
we introduce the {\it first passage time} $\tau_q(W)$, defined as the
(ensemble) average time required by the $q$-th norm of the state vector
${\bf w}(t)$,
\beq
||{\bf w}(t)||_q = \left[ \frac{1}{L} \sum_{i=1}^L w_i^q(t) \right]^{\frac{1}{q}} 
\label{qnorm}
\eeq
to become smaller than some threshold $W$ for the first time. At variance with
Ref.~\cite{ABCPV96,CT01}, we do not care if the evolution of the perturbation
is non monotonous: as we shall see, in this context, the analysis does not only
remain meaningful, but even more, it allows identifying the reason for the
existence of a DP-like scenario even in the context of the continuous model.

At variance with the standard Lyapunov exponent, the FSLE does depend on the
choice of the norm (in particular, on the $q$-value in Eq.~(\ref{qnorm})). This
circumstance is often considered as a difficulty, hindering a proper definition
of FSLE: we prefer to see it as an indication of a richer class of phenomena
associated with the evolution of finite-amplitude perturbations. Anyway, it
has been noticed in Sec.~\ref{sec2} that the ``natural'' order parameter of DP
transition is the spatial average of the state vector. Accordingly, we have
decided, in the present context, to fix $q=1$ (that corresponds to performing an
arithmetic average) and to drop, for the sake of simplicity, the dependence on $q$.

The FSLE $\Lambda(W)$ can be introduced by first fixing a sequence of
decreasing thresholds $W_n$, $n=0,1,2,\ldots$,
\beq
\frac{W_n}{W_{n-1}} = r\quad, \quad\quad r <1 \quad,
\eeq
and by then defining
\beq
\Lambda(W_{n}) = \frac{\ln r}{\tau(W_{n+1})-\tau(W_{n})}
\quad,
\label{FSLE}
\eeq
where  the dependence on the ``discretization''
$\ln r$ is left implicit. In the limit $r \to 1$, the definition becomes
\beq
\Lambda(W)  = 
\left[ \frac{d\,\tau(W)}{d\,(\ln W)} \right]^{-1} \quad .
\label{FSLE2}
\eeq
In the further limit $W \to 0$, $\Lambda(W)$ reduces to the usual
Lyapunov exponent $\lambda$, independently of the adopted $q$-value. When
$\Delta = 0$, $\lambda = \ln a$.

\subsection{Uncoupled limit}
In the uncoupled limit, $\eps =0$, each site converges independently
to the synchronized state (as long as $a<1$). In spite of the low-dimensionality
of the problem, even in this case, an analytic expression for the FSLE can be
obtained only at the expense of introducing further approximations. We shall see
that the resulting expression can be nevertheless profitably used even in the
coupled regime. 

By setting $r=a$ and restricting ourselves to the case $\Delta = 0$, it easy to
show (see appendix \ref{appA}) that
\beq
\tau(W_n) = \frac{\tau(W_{n-1}) + 1}{1-W_n} \quad,
\label{tau0dim}
\eeq    
where $\tau(W_0)=0$, $W_n = a^n$. By inserting Eq.~\toref{tau0dim} in
Eq.~\toref{FSLE} one obtains
\beq
\Lambda(W_n) = \frac{1-a\,W_n}{1+a\,W_n\,\tau(W_n)} \ln a \quad.
\label{0dimFSLE}
\eeq
Eq. \toref{tau0dim} implies that, for $n \to \infty$, 
\beq
\tau(W_n) \sim n + n_0 = \frac{\ln W_n}{\ln a} + n_0 \quad.
\label{approx}
\eeq
By inserting Eq.~\toref{approx} into Eq.~\toref{0dimFSLE} and recalling that
$\lambda = \ln a$ (for $\Delta = 0$), we obtain
\beq
\Lambda(W_n) = \frac{1-a\,W_n}{1+an_0W_n + 
  (a/\lambda)W_n \ln W_n} \lambda \quad .
\label{0dimFSLE1}
\eeq
As we are interested in describing the region where $W_n \ll 1$, and owing
to relative smallness of $a$ ($a\ll n_0$), this equation can be further
simplified to
\beq
\Lambda(W) = \frac{\lambda}{1 +b_0W -b_1W \ln W}  \quad ,
\label{0dimFSLE2}
\eeq
where we have also dropped the unnecessary dependence on the index $n$. 
In the $W \to 0$ limit, the leading correction to the standard Lyapunov
exponent is provided by the term proportional to $b_1$. From the structure of
Eq.~\toref{0dimFSLE2}, it is natural to interpret the inverse of $b_1$ as the
critical threshold $W_c$, below which the dynamics of the uncoupled system
is dominated by the maximum Lyapunov exponent $\lambda$. 

From Eq.~\toref{0dimFSLE2} and by integrating Eq.~\toref{FSLE2}, it is possible to
derive also an analytic expression for the first passage time $\tau(W)$, 
\beq
\tau(W) = \int \Lambda(W) ^{-1} d\, (\ln W) \approx 
\frac{1}{\lambda} \left[ \ln W + (b_0 + b_1) W - b_1 W \ln W \right ] + b_2 \quad,
\label{tauint}
\eeq
where $b_2$ is the
integration constant. In principle, one could imagine of determining $b_2$
by imposing $\tau(1) \equiv b_2 - b_1/\lambda$ equal to 0, since the
evolution starts precisely from $W=1$. However, we cannot expect our
perturbative formula to describe correctly the initial part of the
contraction process. Therefore, $b_2$ must be determined independently.

\subsection{General case and scaling arguments}
In the coupled case, we have not been able to derive an analytical expression
for the FSLE. Nevertheless, a comparison with the numerical results has
revealed that Eqs.~(\ref{0dimFSLE2},\ref{tauint}) describe in a reasonable way
the dependence of $\Lambda$ and $\tau$ on $W$ even in the continuous
model.  However, while $\lambda$ still denotes the standard Lyapunov exponent
of the process and can thus be computed independently, now $b_0$, $b_1$, and
$b_2$ have to be determined by fitting the numerical data. We have preferred
to keep also the term proportional to $b_0$ (relevant only for relatively large
$W$-values), since its presence guarantees a much better reproduction of
the numerical data. In Fig.~\ref{passage-times-fit}a, we see that
Eq.~\toref{tauint} provides a good parametrization of the numerically
determined $\tau$-values over a wide range of thresholds, both in the
discontinuous and the continuous model (see the solid curves). In panel b) we
notice that, although the theoretical expression \toref{0dimFSLE2} does not
provide an equally good description of the FSLE, it is nevertheless able to
pinpoint the crossover towards the small-amplitude behaviour of the
perturbation. We will see that the possibility of identifing the largest scale
(defined by $1/b_1$) over which the linearized dynamics (described by the
standard Lyapunov exponent) sets in represents a crucial point of our analysis.

\begin{figure}[tcb]
\centering
\includegraphics[width=10cm, angle=0]{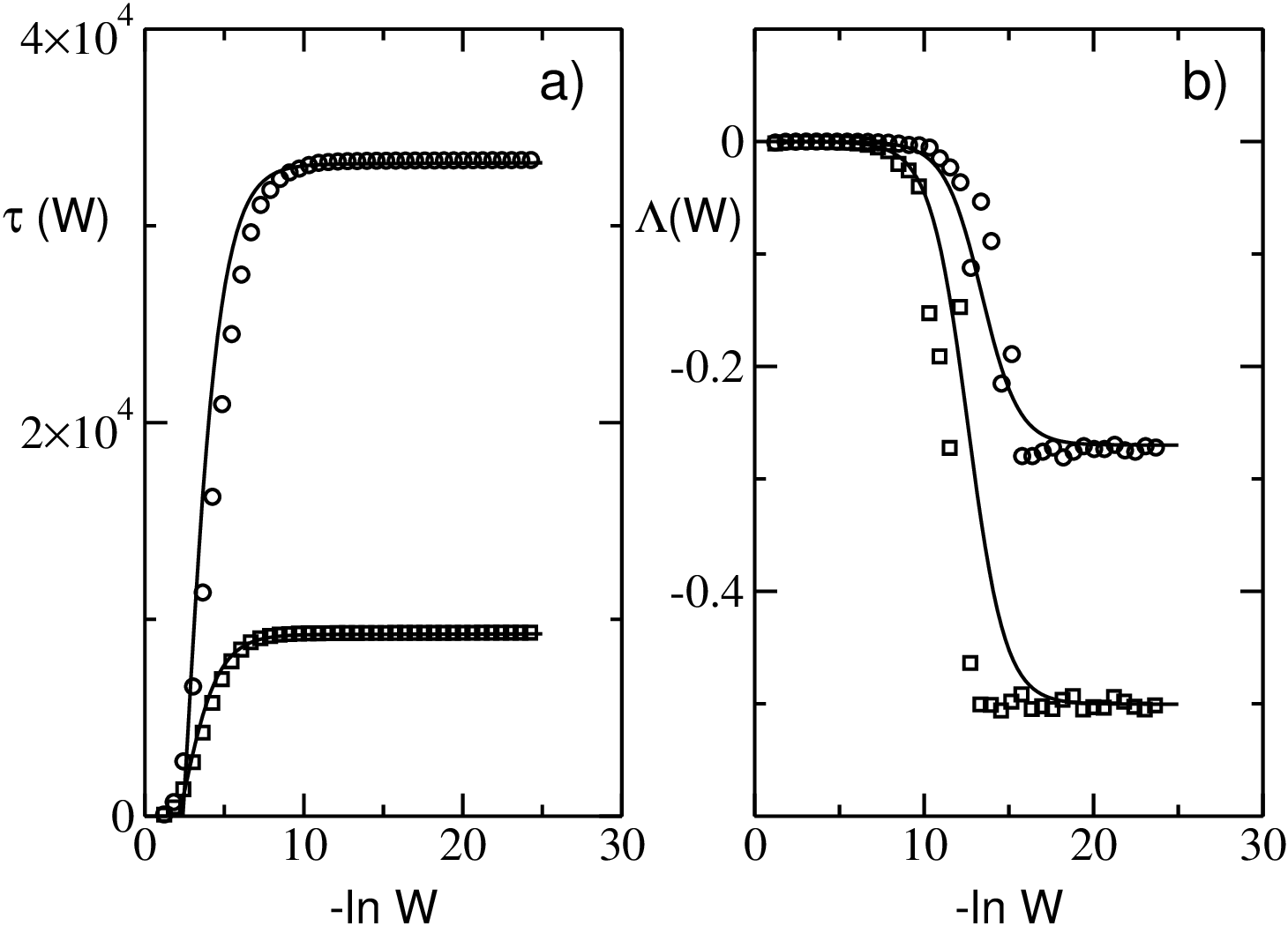}  
\caption{Numerical data for the first passage time (left panel) and the
FSLE (right panel) at the critical point 
are fitted with Eq.~\toref{tauint} and Eq.~\toref{FSLE2} (full lines). 
Circles refers to $\Delta=0.01$, $L=256$, $a_c = 0.6055$ and squares to
$\Delta=0$, $L=128$, $a_c=0.6063$.}
\label{passage-times-fit} 
\end{figure} 

It is now natural to ask to what extent Eq.~(\ref{tauint}) is able to account
for the scaling behaviour in the vicinity of the transition. By replacing
$\rho$ with $W$ in Eq.~(\ref{scale1a}) and $t$ with the first passage
time $\tau$, one expects that, at criticality,
\beq
 W = L^{-\delta z} g(t/L^z) \quad.
\label{scale1}
\eeq
Inversion of this equation leads to
\beq
\tau(W) = L^{z} g^{-1}(W L^{\delta z})\quad.
\label{scale2}
\eeq
Before mutually comparing the two expressions (\ref{tauint},\ref{scale2}), it
should be first stressed that they have been introduced to address different
questions. On the one hand, Eq.~(\ref{tauint}) is an approximate expression
introduced to account for the crossover towards the $W$-range where
the dynamics is controlled by linear mechanisms and no scaling behavioiur should
be expected. On the other hand, Eq.~(\ref{scale2}) is a rigorous but implicit
statement about the scaling region only.

Compatibility between Eqs.~(\ref{tauint}) and (\ref{scale2}) requires a proper
dependence of $b_0$, $b_1$ and $b_2$ on the systems size $L$, namely 
\begin{eqnarray}
b_0 &=&  -\lambda \left[\tilde{b}_0-\tilde{b}_1(1+\delta z \ln L)\right] L^{z(1+\delta)} \\
b_1 &=& -\lambda \tilde{b}_1 L^{z(1+\delta)}\\
b_2 &=& \tilde{b}_2 L^z 
\quad ,
\label{btau}
\end{eqnarray}
where $\tilde{b}_0$, $\tilde{b}_1$, and $\tilde{b}_2$ are 
suitable positive constants. By inserting Eq.~\toref{btau} into Eq.~\toref{tauint}, 
one finds that
\beq
\tau(W) = \frac{\ln W}{\lambda} - L^z \left[\tilde{b}_0 W L^{z\delta} -
  \tilde{b}_1 W L^{z\delta}\ln (W L^{\delta z}) + \tilde{b}_2 \right ]\quad,
\label{tauscaled}
\eeq
from which we see that the first term in the r.h.s. is the only one which does
not follow the required scaling law (\ref{scale2}). In fact, $(\ln
W)/\lambda$, accounts for the linearly stable behaviour in a regime where
a finite-state model (such as, e.g., the famous Domany-Kinzel model \cite{DK84})
would be otherwise characterized by a perfect absorption (when a configuration
of all 0's is attained).

In order to test the correctness of the whole picture, we have studied the
dependence of $b_1$ and $b_2$ on $L$. In Fig.~\ref{passage-times-scaling}, 
their behaviour is plotted at criticality for the discontinuous and the
continuous model: both quantities show a good agreement with the power law
divergence predicted by Eq.~\toref{btau} ($z\approx 1.58$ and $z(1+\delta)
\approx 1.82$). As for the last parameter $b_0$, given its involved dependence
on $L$ and the approximate character of Eq.~(\ref{tauscaled}), we can only claim
that its dependence is qualitatively consistent with the theoretical prediction.

\begin{figure}[tcb]
\centering
\includegraphics[width=10cm, angle=0]{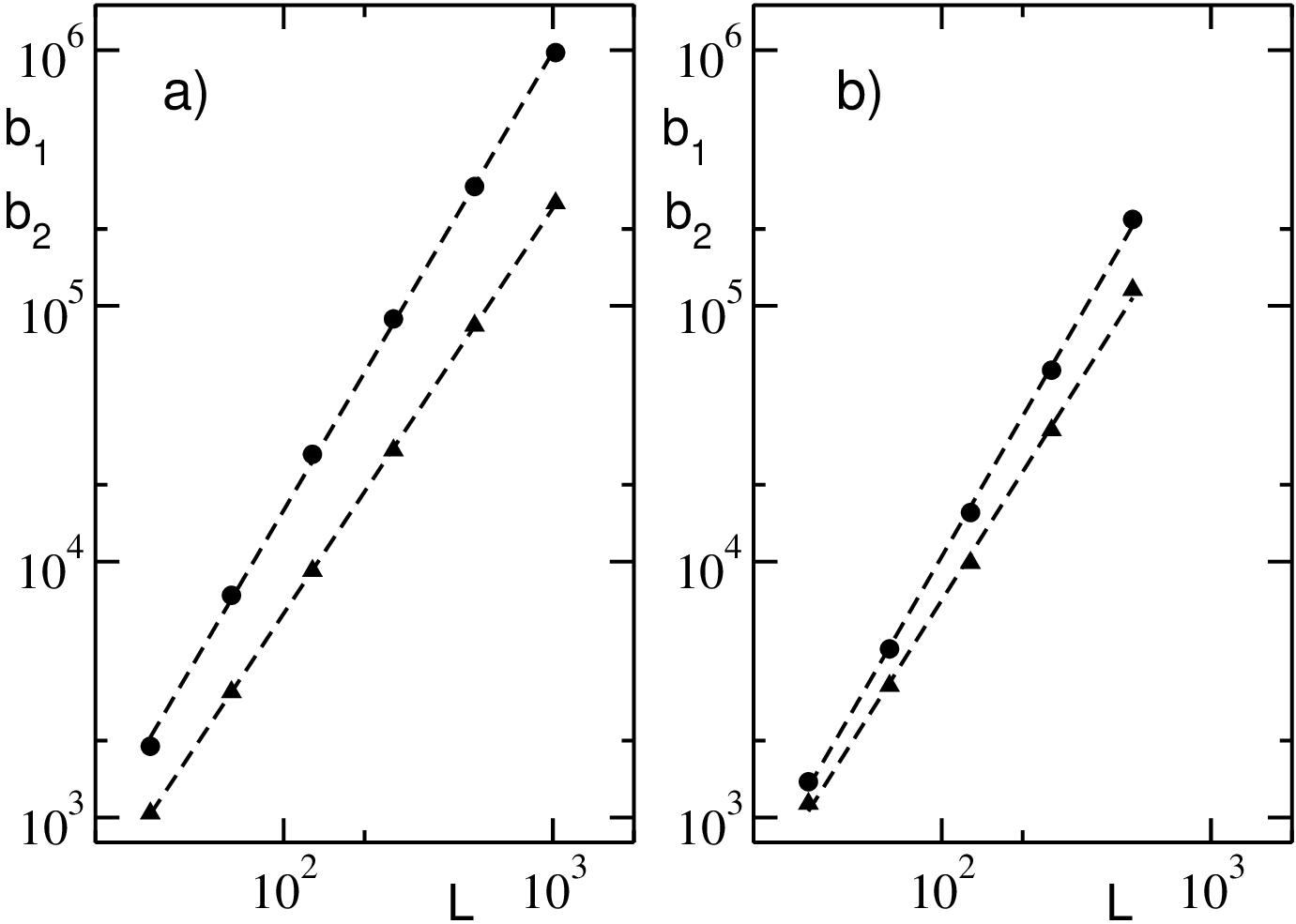}  
\caption{Finite-size scaling behaviour at the critical point   of both $b_1$
and $b_2$ for $\Delta=0$ and $a_c=0.6063$ (left panel)  and $\Delta=0.01$
and $a_c=0.6055$ (right panel). The dashed lines indicate the best  power law
fit.  Left: $b_1$ (circles) scales as $L^{1.79}$, while $b_2$ (triangles)
scales as $L^{1.58}$.  Right: $b_1$ (squares) scales as $L^{1.82}$, while
$b_2$ (diamonds) scales as $L^{1.66}$.}
\label{passage-times-scaling} 
\end{figure} 
One of the most important results of our study is the objective identification
of a threshold $W_c = 1/|b_1|$, below which linear stability analysis
holds and its scaling dependence on $L$ ($W_c \sim  L^{-z}$). In a model like
the cellular automaton considered by Domany and
Kinzel \cite{DK84}, absorption in a finite system occurs when all sites collapse
onto the absorbing state: this means that the minimal meaningful density that
can be considered is $\rho_m = 1/L$. In the present context, $W_c$ plays
the role of $\rho_m$: below $W_c$, the critical behaviour is dominated
by the linearly stable dynamics. The difference between the two systems lies
in the scaling dependence of the maximal resolution on $L$. Since $W_c$
decreases faster than $1/L$ this means that, e.g., the scaling range for
$W_c$ is wider in the present model than in finite-state systems.

Finally, we comment about the reason why the range of validity of the linear
stability analysis can eventually vanish even in models like the continuous RM,
where every perturbation locally smaller than $\Delta$ should behave linearly.
The reason is that $\tau(W)$ is defined as the average first-passage time:
even if the perturbation is homogeneously small, if $L$ is sufficiently large,
some occasional amplification may occur and drive, on the average, the system
out of the linear region. It is only below $W_c$ that such sporadic
resurgencies are sufficiently rare not to modify significantly the stable
linear behaviour.

\section{Conclusions and open problems}
\label{sec5}

In this paper we have expounded a partially rigorous argument to show why the
synchronization transition in spatially extended systems may belong to the DP
universality class. Although our theoretical considerations restrict to the
discontinuous RM model, scaling analysis of the first-passage time
$\tau(W)$ suggests that the transition belongs to the DP class also in a
finite parameter region of the continuous model. Since direct numerical
simulations in the more physical class of CMLs have been basically restricted
to discontinuous maps, we find it wise to test the validity of our conclusions
also in the context of continuous, though highly-nonlinear maps. Accordingly,
we have considered two lattices of maps coupled as in Eq. \toref{chaosync}; 
the local map is chosen similar to those defined by
Eq.~(\ref{detmap}), namely
\beqn 
f(x)= \left\{  \begin{array}{lr} 
              x/\alpha_1  & 0\leq x < \alpha_1 \\ 
             1-(x-\alpha_1)(1-\alpha_3)/\alpha_2 &  
	     \quad \quad \alpha_1 \leq x < \alpha_1+\alpha_2 \\
             \alpha_3+\alpha_4(x - \alpha_1-\alpha_2) &  \alpha_1+\alpha_2 \leq x \leq 1  
\end{array} 
\right.\quad . 
\label{detmap2} 
\eeqn 
with $\alpha_1=1/2.7$, $\alpha_3=0.07$, $\alpha_4=0.1$.  The reason for this
choice is that in Ref.~\cite{PT94} it has been shown that in such a model (for
the same parameter values and  $\alpha_2 < 0.013$ \cite{foot3}) nonlinear
effects prevail over linear ones. In fact, it was observed that the propagation
velocity $v_F$ of finite-amplitude perturbations (see Eq.~(\ref{vF})) is larger
than the propagation velocity $v_L$ of infinitesimal perturbation (for a
definition of $v_L$, see \cite{PT92,LPT96}). For instance,
for $\alpha_2=4 \cdot 10^{-4}$ and $\eps = 2/3$, $v_L = 0.4184$,
while $v_F = 0.5805$. In this regime,
upon varying the coupling strength $\sigma$, synchronization arises through a
continuous phase transition accompanied by a negative transverse Lyapunov
exponent and a vanishing $v_F$ at the critical point $\sigma_c =
0.17756\ldots$. As it can be appreciated in Fig.~\ref{chaosync1}, where we have
plotted the space averaged difference variable  $w(t)$ versus
time for different values of the control parameter, the critical decay rate
is $\delta = 0.158 \pm 0.01$, fully compatible with the expectation for a DP
transition. We are thus reinforced in the conjecture that the DP scenario is
robust and not just restricted to the highly nongeneric case of discontinuous
maps.

\begin{figure}[tcb]
\centering
\includegraphics[width=8cm, angle=0]{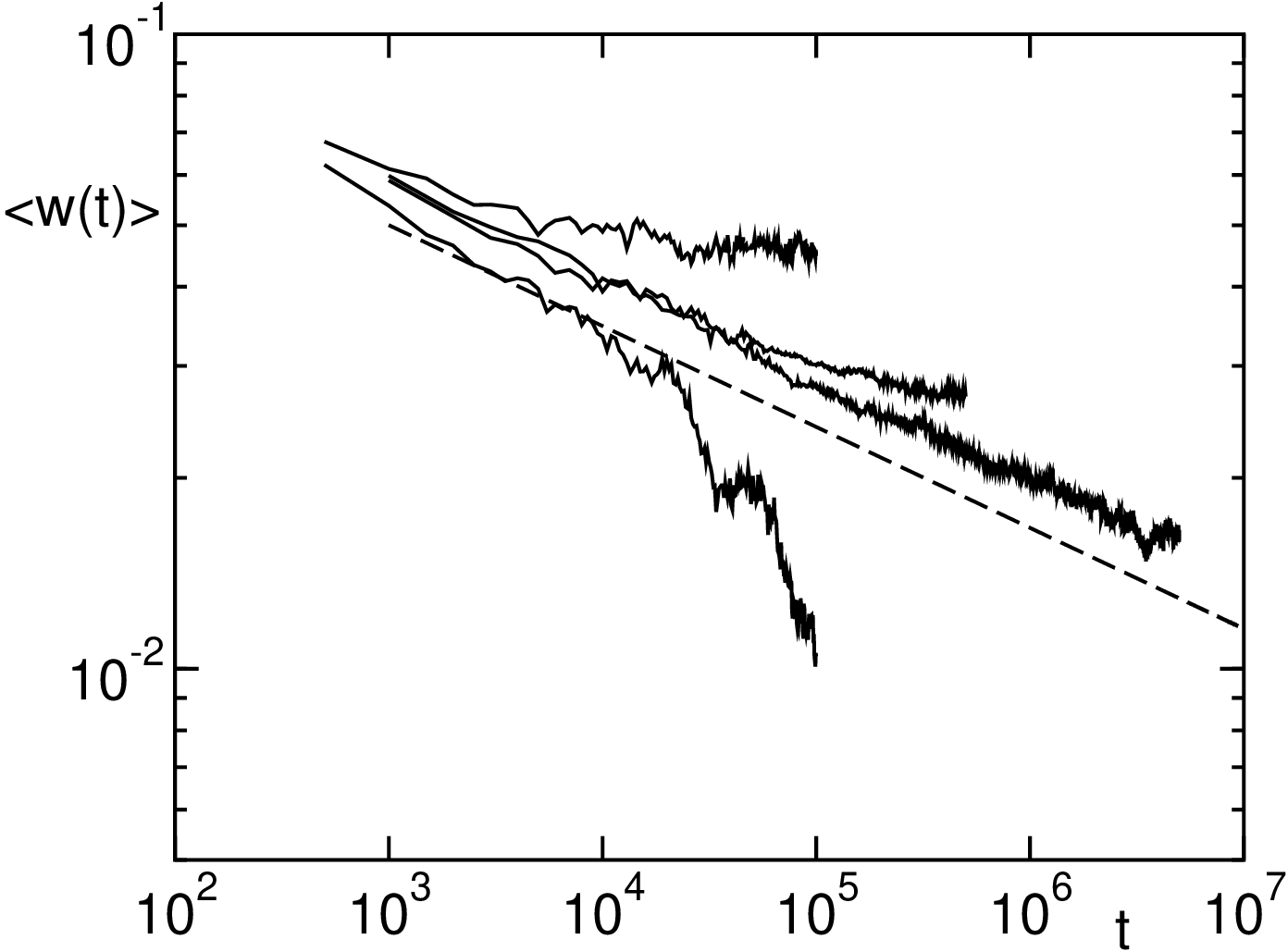}
\caption{Log-log plot of the space averaged difference variable $w(t)$ 
as a function of
time for two coupled CML's (see text) and for different coupling
values: from  lower to upper, the full lines correspond to $\sigma=0.178$,
$\sigma=0.17756$, $\sigma=0.1775$, $\sigma=0.177$, while  the long dashed line
marks the power low expected for the DP critical behaviour. Numerical data have
been obtained averaging over $100$ realizations of a CML of size $L=2^{17}$.}
\label{chaosync1}
\end{figure}
The crucial difficulty to determine the universality class for the
synchronization transition is that the order parameter (the difference field)
can be arbitrarily small. This casts doubts about the very definition of the
zero-difference field as a truly absorbing state. In fact, in a previous paper
\cite{GLP02}, it was speculated that the DP scaling behaviour might be restricted
to a finite range. The analysis carried on in this paper clarifies that the
synchronization transition genuinely belongs to DP universality class: this has
been understood from an objective identification of the threshold $W_c$
below which the dynamics is really controlled by linear mechanisms and thus
corresponds to an effective contraction. The parametrization of $\tau(W)$
introduced to describe the single-map case has greatly helped to unveil the
overall scenario since it has clarified that the basic effect of the diffusive
coupling is to renormalize the parameters defining $\tau(W)$ (see
Eq.~(\ref{tauint})). Here, the parameter values (in particular $W_c$) have
been inferred by fitting the numerical data; in the future, it will be
desirable to find an analytic, though approximate, way of performing the
renormalization. 

Once we have concluded that synchronization arises through a DP-like transition
in a finite parameter region, it is natural to ask how this scenario crosses
over to the standard transition characterized by a vanishing of the
Lyapunov exponent and by the KPZ critical exponents. With reference to
Fig.~\ref{RMPhase}, this question amounts to investigating the region around the
multicritical point $\Delta_c$. A purely numerical analysis of this region
is not feasible in this model, as it would require considering too large
systems to be effectively handled. We are currently studying this problem in a
different context, where preliminary studies indicate the possibility to draw
quantitative conclusions. 

Finally, since it is known that finite-size Lyapunov exponents do depend on the
norm, it might be worth considering $q$ values different from 1, in order to
check to what extent the universality of the transition is preserved when
different averaging procedures are adopted to assess the amplitude of the
global perturbation. In particular, since $q = \infty$ (corresponding to the
maximum norm) takes care only of the extreme fluctuations of a perturbation
field, it is not totally obvious that the behaviour of the corresponding first
passage time follows exactly the above described scenario.

\begin{acknowledgments}
We thank A. Pikovsky, V. Alhers, P. Grassberger and D. Mukamel for fruitful
discussions and suggestions. CINECA in Bologna and INFM are acknowledged for
providing us access to the parallel Cray T3 computer through the grant
``Iniziativa calcolo parallelo''.
\end{acknowledgments}

\appendix
\section{First passage times in the discontinuous uncoupled RM model}
\label{appA}
In this appendix we report the analytical calculation of the first passage time
when $\eps=0$ and $\Delta=0$, to prove Eq. \toref{tau0dim}. 
Being $W_n = a^n$ and $w(0)=1$, we have also $\tau(W_0) = 0$. In order to 
compute the first passage time through a threshold $W_n$, we need to know
the average time needed to pass from $W_{n-1}$ to $W_n$.  With a probability 
$1 - a W_{n-1}$, this can occur in one time step, if the amplification 
mechanism is not activated and the synchronization error is contracted by a 
factor $a$. On the other hand, with probability $a W_{n-1}$, the amplification 
resets the state variable to the value 1. In this case, one has to wait 
for the synchronization error to shrink back to the $n$-th threshold, 
which, by definition, occurs in an average time $\tau_{n-1}$. 
At this point, the error can either shrink to $W_n$ or be reset again to 
1, to start again the process. Altogether, 
\begin{eqnarray}
\tau(W_n) = & \tau(W_{n-1}) + 1 \cdot (1 - aW_{n-1}) + 
(2 + \tau(W_{n-1}) (1 - aW_{n-1}) a W_{n-1} + \nonumber \\
& (3 + 2\tau(W_{n-1}) (1 - aW_{n-1}) (a W_{n-1})^2 + \ldots \nonumber \\ 
& = \tau(W_{n-1}) + (1 - aW_{n-1})  \sum_{i=0}^{\infty} 
\left[ (1+ i + i \tau(W_{n-1}) ) (a W_{n-1})^i  \right ] \nonumber \\
& = \tau(W_{n-1}) + (1 - a^n) \left [ \sum_{i=0}^{\infty} (a^n)^i 
 + (\tau(W_{n-1} + 1) \sum_{i=0}^{\infty} i\,(a^n)^i \right ] \quad.
\end{eqnarray}
Summing up the series, one obtains 
\beq
\tau(W_n) = \frac{\tau(W_{n-1}) + 1}{1-a^n} \quad.
\eeq



\begin{thebibliography}{99}
\bibitem{foot1} Having in mind to work with discrete-time systems, it is
natural to adopt the It\^o interpretation.
\bibitem{foot2} This is indeed true for coupled Bernoulli maps 
$F(x) = rx \;\mod 1  $, even though the diffusive coupling induces some 
inhomogeneity in the flat distribution of the single map.
\bibitem{foot3} Notice that $\alpha_2$ plays here the same role as 
$\Delta$ in the RM.
\end{thebibliography}
\end{document}